\documentclass[a4paper, aps, superscriptaddress, amsfonts, amssymb, amsmath, reprint, showkeys, nofootinbib]{revtex4-2}
\usepackage[english]{babel}
\usepackage[utf8]{inputenc}
\usepackage[colorinlistoftodos, color=green!40, prependcaption]{todonotes}
\graphicspath{{}}

\begin{document}

\title{Quantum Hall effect in lightly hydrogenated graphene}

\author{I.G.~van Rens}
\email[Correspondence email address: ]{inge.vanrens@ru.nl}
\affiliation{Institute for Mathematics, Astrophysics and Particle Physics (IMAPP), Radboud University and Nikhef, Heyendaalseweg 135, 6525 AJ Nijmegen, The Netherlands}
\affiliation{HFML-FELIX / EMFL, Toernooiveld 7, 6525 ED Nijmegen, The Netherlands}
\author{O.O.~Zheliuk}
\affiliation{HFML-FELIX / EMFL, Toernooiveld 7, 6525 ED Nijmegen, The Netherlands}
\affiliation{Institute for Molecules and Materials, Radboud University, Heyendaalseweg 135, 6525 AJ Nijmegen, The Netherlands}
\author{M.W.~de Dreu}
\affiliation{HFML-FELIX / EMFL, Toernooiveld 7, 6525 ED Nijmegen, The Netherlands}
\affiliation{Institute for Molecules and Materials, Radboud University, Heyendaalseweg 135, 6525 AJ Nijmegen, The Netherlands}
\author{K.~Mukhuti}
\affiliation{HFML-FELIX / EMFL, Toernooiveld 7, 6525 ED Nijmegen, The Netherlands}
\affiliation{Institute for Molecules and Materials, Radboud University, Heyendaalseweg 135, 6525 AJ Nijmegen, The Netherlands}
\author{Y.~Kreminska}
\affiliation{Device Physics of Complex Materials, Zernike Institute for Advanced Materials, University of Groningen, Groningen, The Netherlands}
\author{{C.S.A.~M{\"u}ller}}
\affiliation{MESA+ Institute for Nanotechnology, University of Twente, P.O. Box 217, Enschede, 7500 AE The Netherlands}
\author{P.C.M.~Christianen}
\affiliation{HFML-FELIX / EMFL, Toernooiveld 7, 6525 ED Nijmegen, The Netherlands}
\affiliation{Institute for Molecules and Materials, Radboud University, Heyendaalseweg 135, 6525 AJ Nijmegen, The Netherlands}
\author{J.T.~Ye}
\affiliation{Device Physics of Complex Materials, Zernike Institute for Advanced Materials, University of Groningen, Groningen, The Netherlands}

\author{N.~de Groot}
\affiliation{Institute for Mathematics, Astrophysics and Particle Physics (IMAPP), Radboud University and Nikhef, Heyendaalseweg 135, 6525 AJ Nijmegen, The Netherlands}
\author{U. Zeitler}
\email[Correspondence email address: ]{uli.zeitler@ru.nl}
\affiliation{HFML-FELIX / EMFL, Toernooiveld 7, 6525 ED Nijmegen, The Netherlands}
\affiliation{Institute for Molecules and Materials, Radboud University, Heyendaalseweg 135, 6525 AJ Nijmegen, The Netherlands}

\date{\today} 

\begin{abstract}
We have measured the quantum Hall effect in monolayer graphene samples that were exposed to a cold hydrogen plasma leading to a hydrogenation level of a few percent. Compared to pristine graphene, the Landau level distance significantly decreases in the hydrogenated structures, and its field dependence changes from square root type to linear. From this observation we conclude that the band structure in hydrogenated graphene changes from a linear Dirac-Weyl type dispersion to a quadratic one with an effective electron mass  $m_e^* = 0.24~m_e$. This is in good agreement with ab-initio band structure calculations of hydrogen decorated graphene monolayer.
\end{abstract}

\maketitle

\section{Introduction}

In the Standard Model, neutrinos do not couple to the Higgs field and are therefore massless \cite{neutrino_mass}. Massive neutrinos would constitute evidence for physics beyond the Standard Model.  Neutrino oscillation experiments have established that neutrinos possess nonzero masses \cite{KAJITA1999123}. Oscillation experiments do not measure the absolute masses of neutrinos; they are sensitive only to differences between the squares of the mass eigenvalues, such as $\Delta m_{12}$ or $\Delta m_{31}$. The absolute neutrino mass scale is therefore still unknown. Its very small magnitude—at least seven orders of magnitude lower than that of the electron, the lightest charged fermion—raises important theoretical questions. The determination of the absolute scale of neutrino masses remains an open problem in contemporary physics, relevant to particle physics, nuclear physics, and cosmology.

The measurement of the neutrino mass through the precise analysis of beta-decay spectra of tritium, is the most direct experimental method available today. Currently, the upper limit on the neutrino mass is set by KATRIN \cite{aker2022katrin} and approaches the limit of what can be measured using molecular tritium (T$_2$). To further lower this limit, and hopefully to measure the neutrino mass more accurately, the Ptolemy collaboration \cite{bettsdevelopment, betti2024cosmic} proposes to instead measure the energy spectrum of beta decay of tritium atoms on
a graphene substrate with the prospect of improving the energy resolution limited by rotations and vibrations of the T$_2$ molecules.

In this paper, we study the viability of graphene \cite{novoselov2004electric, geim2007rise} as a substrate for tritium in the Ptolemy experiment with a first objective of addressing the electronic properties of tritiated graphene, and, in particular, its band structure. Direct experiments with tritium are complex and hazardous \cite{fairlie2008hazards} and, to our knowledge, only a few papers addressing the electronic and optical properties of tritiated graphene experimentally have been published recently \cite{D3NA00904A, becker2025graphenestructuremodificationtritium, zeller2025followinglongtermevolutionsp3type}.

Therefore, in our present work we have first investigated the band structure of hydrogenated graphene, which is supposed to be an isostructural and isoelectronic analogue of tritiated graphene and is more abundant and easier to work with. 
More specifically, we have measured the quantum Hall effect in hydrogenated graphene in a magnetic field up to 30 T and we find that the electron and hole dispersion change from a linear to a quadratic one. From the field dependence of the Landau-level gaps we determine that the effective cyclotron mass in our hydrogenated samples was m$^*$ = 0.24 $\pm$ 0.04 m$_{\mathrm{e}}$ and  m$^*$ = 0.4 $\pm$ 0.1 m$_{\mathrm{e}}$  which is in reasonable agreement with simple band-structure simulations of hydrogenated graphene with 4 $\pm$ 1\% and 7 $\pm$ 2\% hydrogen coverage, respectively.

\begin{figure*}[t]
    \centering
    \includegraphics[width=\linewidth]{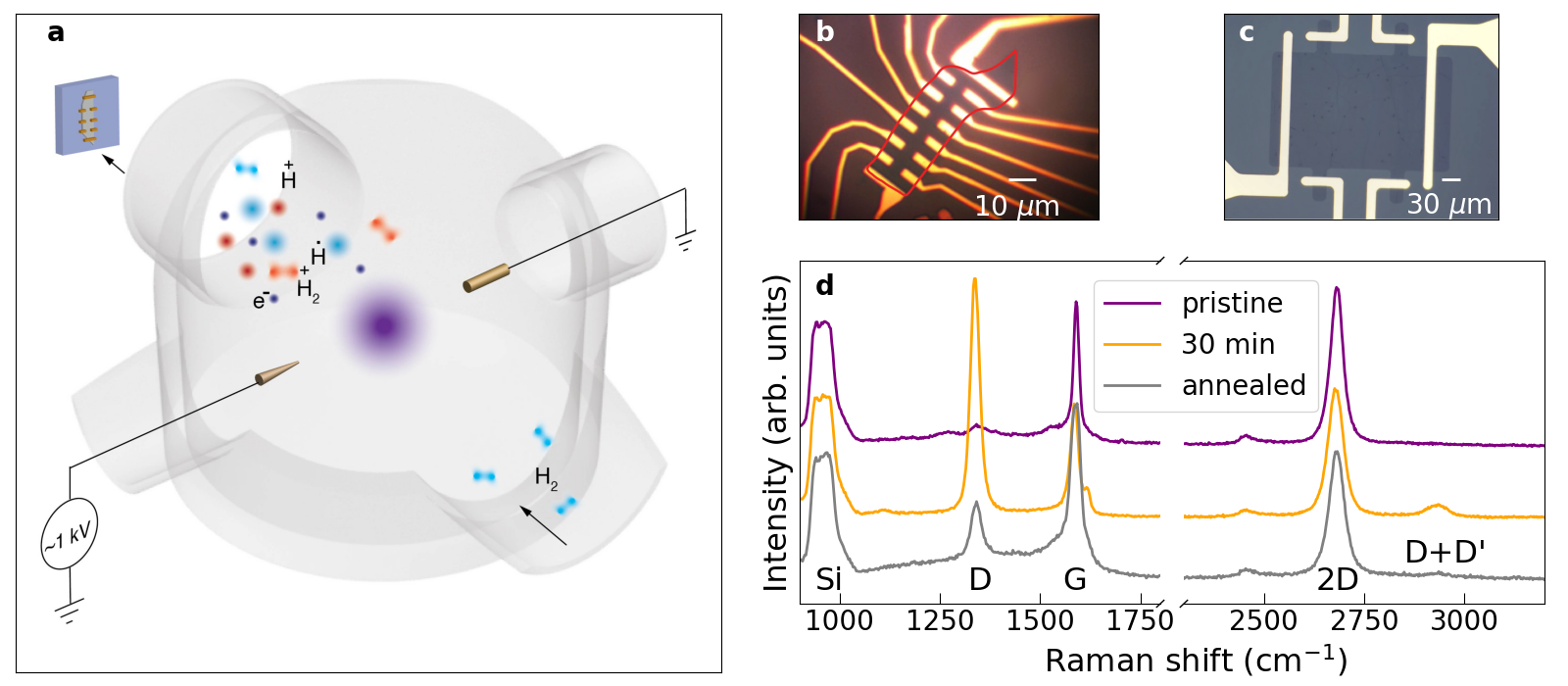}
    \caption{(a) Cold plasma hydrogenation setup. The graphene sample is located in the top left corner. The plasma is located in the middle. Hydrogen gas flows in from the bottom right. (b) Exfoliated graphene sample. The channel width is 5 $\mu$m. The red contour shows the graphene flake. (c) CVD graphene sample acquired from graphenea \cite{graphenea}. The graphene flake is visible as the darker blue area. (d) Raman spectra of the CVD graphene sample in consecutive states: pristine, after 30 minutes of hydrogen plasma exposure, and after annealing at 300 $^\circ$C. Peaks correlated to graphene peaks and graphene defect peaks are indicated. }
    \label{fig:hydrogenation}
\end{figure*}

\section{Experimental Methods}

Two types of graphene devices were used in this study. The first type was prepared by standard mechanical exfoliation of graphite onto an $n$-doped Si wafer (acting as a back-gate) with 295 nm of SiO$_2$ as gate insulator on top of it (Fig.~\ref{fig:hydrogenation}b). Monolayer flakes of typical dimensions 5-10 $\mu$m wide and 20-40 $\mu$m long were selected and electrode regions for ohmic contacts were defined by e-beam lithography followed by Ti/Au metal deposition of 5/50 nm in thickness. The second type were commercially available samples \cite{graphenea} made with chemical vapor deposition (Fig.~\ref{fig:hydrogenation}c). The substrate was again SiO$_2$/Si+ with a wafer thickness of 675 $\mu$m and an oxide thickness of 90 nm. These samples had Au contacts in a standard Hall-bar layout.

In order to (partially) hydrogenate our graphene sample, we used a cold-plasma hydrogenation method \cite{elias2009control} as shown in Fig.~\ref{fig:hydrogenation}a. A constant flow of hydrogen gas is fed through a plasma chamber at 1 mbar. A voltage of 1 kV is applied between two closely spaced Al electrodes to create a plasma, where hydrogen radicals diffuse slowly towards the sample located 20 cm from the plasma region, and some bond to the graphene \cite{ZHAO2021244}. It should be noted that additional effects such as etching of the graphene at low pressure \cite{felten2014insight, despiau2016hydrogen}, can occur.
After performing experiments on the hydrogenated graphene samples, we have annealed some of them at 300 $^\circ$C in a 500 mbar Ar atmosphere to move back towards the pristine state.

To further characterize their structural properties, we have also characterized some of our graphene samples using Raman spectroscopy with a 532 nm (2.33 eV) laser at 225 $\mu$W laser power focused to a spot size with a diameter of several microns. For each state, 3 measurements were taken with a five minutes accumulation time. The results of this measurement performed on a CVD-grown sample are shown in Fig.~\ref{fig:hydrogenation}d. The pristine sample shows the expected spectrum, with a G peak at ~1580 cm$^{-1}$ and a 2D peak at ~2670 cm$^{-1}$. After hydrogen plasma exposure, the G and 2D peak become smaller. An additional defect-induced D peak appears at ~1340 cm$^{-1}$, along with a D' peak at ~1620 cm$^{-1}$ on the left shoulder of the G peak, and a D+D' peak at ~2920 cm$^{-1}$. These peaks match literature on hydrogenated graphene \cite{elias2009control, luo2011electronic}. After annealing the effects of hydrogenation are partially reverted, which confirms that the changes are mainly caused by addition of hydrogen \cite{wojtaszek2011road}.

\begin{figure}
    \centering
    \includegraphics[width=\linewidth]{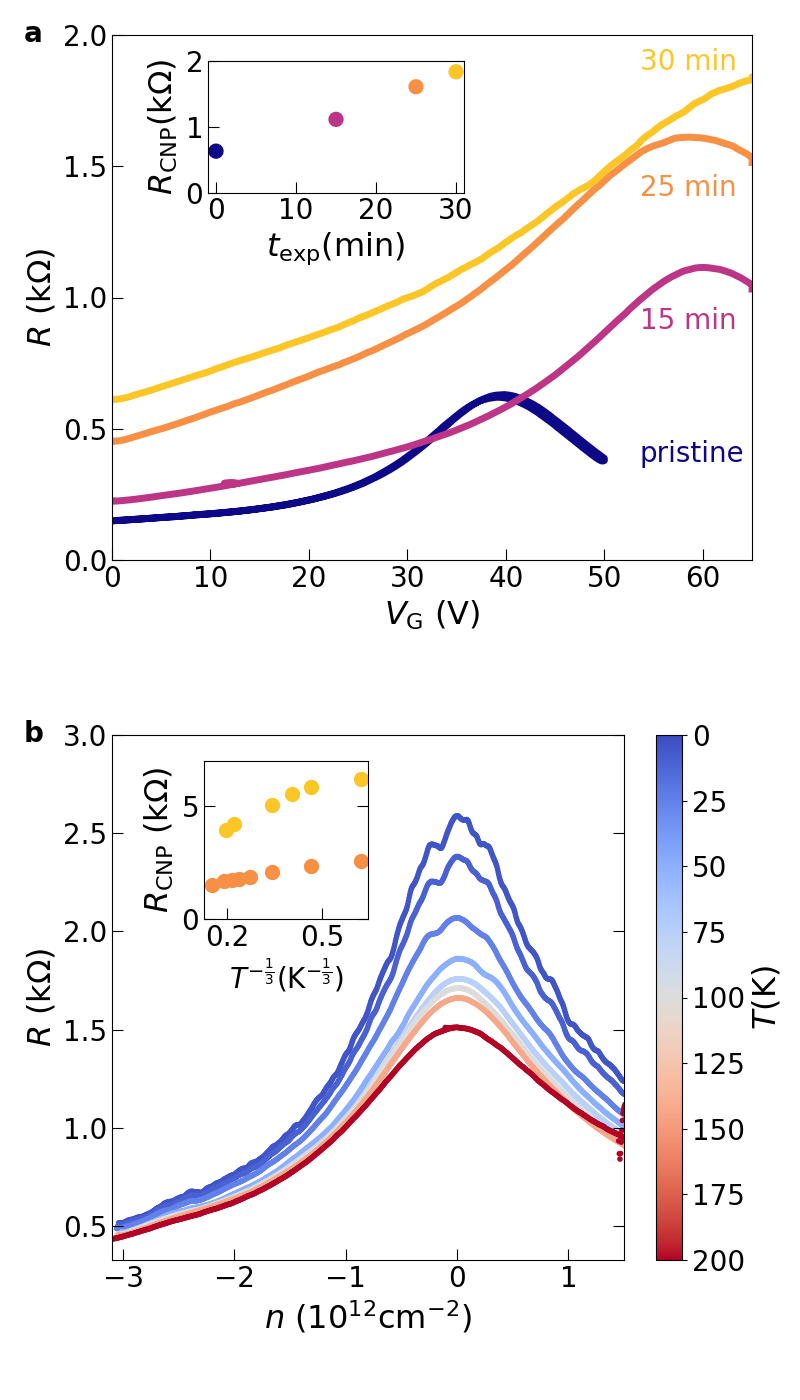}
    \caption{a.~Change of the transfer curve at room temperature as a function of hydrogen plasma exposure, measured on an exfoliated graphene sample. Color coding matches the points in the inset. Inset: room temperature CNP resistance as a function of hydrogen plasma exposure time. b.~Temperature dependence of the transfer curve at 0 T for the exfoliated graphene sample shown in Fig.~\ref{fig:transport}a, exposed to hydrogen plasma for 25 minutes. The temperature is encoded in the color of the lines (right color bar).  Inset: temperature dependence of the CNP resistance for a cleaved graphene sample that was exposed to hydrogen plasma for 25 (orange) and 30 minutes (yellow). }
    \label{fig:transport}
\end{figure}

\section{Experimental Results and Discussion}
\subsection{Transfer curve of hydrogenated graphene}

We now move to a further characterization of our samples by studying the sample resistance as a function of gate voltage for different levels of hydrogen plasma exposure (Fig.~\ref{fig:transport}a). After plasma exposure, the charge neutrality point (CNP) shifts to higher gate voltage due to \textit{p}-type doping. This can be explained by physisorption of molecules such as H$_2$O \cite{schedin2007detection} or by bonding H$^+$ ions onto the graphene. We also can notice a resistance increase at the CNP and a broadening of the peak caused by the addition of extra defects decreasing the charge carrier mobility. This is a typical behaviour for all of our hydrogenated graphene samples, both CVD-grown and exfoliated, with variations roughly proportional to the hydrogen exposure time. However, more complex effects can influence the hydrogenation process and the time alone is not always a good measure for the actual hydrogen coverage. At higher plasma exposure a band gap starts to open, causing the resistance to increase further and to become temperature dependent as shown in Fig.~\ref{fig:transport}b  after 25 minutes of hydrogen plasma exposure. The inset shows the CNP resistance after 25 (orange) and 30 (yellow) minutes of plasma exposure. Both the resistance and the temperature dependence keep increasing with additional plasma exposure. 

\begin{figure*}
    \centering
    \includegraphics[width=\linewidth]{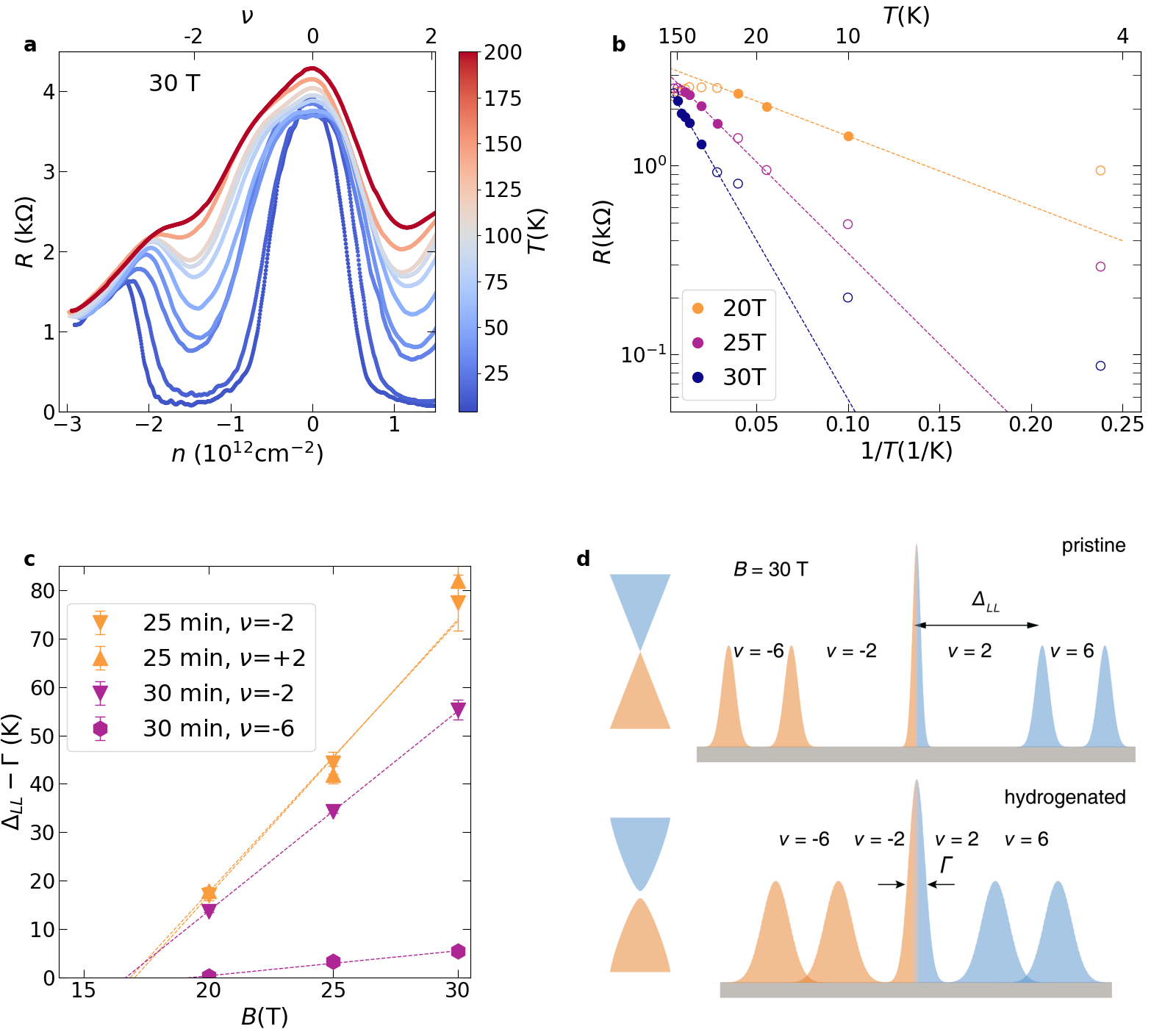}
    \caption{a. Temperature dependence of hydrogenated graphene shown in Fig.~\ref{fig:transport}b as a function of charge carrier concentration (bottom axis) and filling factor (top axis) at 30 T. The temperature is encoded in the color of the lines (right color bar). Resistance minima are visible at filling factors $\pm$2. b. Fit of the resistance minima for $\nu=-2$ visible in Fig.~\ref{fig:QHE}a to thermal activation. c. Landau-level activation gaps as a function of magnetic field for two hydrogenation states of the same sample. d. Density of states in magnetic field for pristine and hydrogenated graphene. After hydrogenation the Landau levels broaden and the gaps between them decrease. This figure is not to scale.}
    \label{fig:QHE}
\end{figure*}

\subsection{Landau levels and excitation gaps}

In order to further characterize and understand the electronic properties of hydrogenated graphene we have performed magnetotransport experiments in fields up to 30 T. This allows to resolve the Landau-level structure and, from the energetic difference between Landau levels, extract the cyclotron mass of the charge carriers. In Fig.~\ref{fig:QHE}a we show the resistance of a hydrogenated graphene sample in a magnetic field of 30 T for temperatures varying from 4 K to 200 K as a function of charge carrier concentration. These experiments were performed on a exfoliated graphene sample that was exposed to a hydrogen plasma for 25 minutes. The resistance minima at filling factors $\pm$2 as known from pristine graphene are still clearly visible in the hydrogenated graphene, but start vanishing at comparably lower temperature. This temperature dependence is shown in Fig.~\ref{fig:QHE}b in the form of an Arrhenius plot of the resistance minima. From the slope of these plots the activation gap can be extracted using
\begin{equation}
    R_{\mathrm{min}} = e^{-\frac{\Delta_a}{k_BT}}.
\end{equation}

Here, $R_{\mathrm{min}}$ is the resistance at the minimum, $k_\mathrm{B}$ the Boltzman constant and $\Delta_\mathrm{a}$ is the activation gap defined as the distance between the Fermi energy situated in the middle between two Landau levels and the conductivity edge of the adjacent Landau levels. The linear fit shown in the figure was made to the filled points in the plot. At lower temperatures, the temperature dependence of the resistance minima started to flatten of. These points were excluded from the fit.

Fig.~\ref{fig:QHE}c shows the quantum-Hall activation gaps of the exfoliated graphene sample as a function of magnetic field for several filling factors after 25 and 30 minutes of hydrogen plasma exposure. Comparing these data to literature values on pristine graphene, such as in \cite{GIESBERS20081089} and \cite{giesbers2007quantum}, we find that  the activation gaps of hydrogenated graphene are far smaller than in pristine graphene: at 30 T, the activation gap for $\nu$=$\pm$2 in pristine graphene is $\sim$2000 K, while we measure a gap of 80 K after 25 minutes of hydrogen plasma exposure. After 30 minutes of exposure this gap shrinks even further to 55 K. Furthermore, the minimal field necessary to see the quantum Hall effect has increased from typically 5 T \cite{giesbers2007quantum} to 20 T which can be interpreted as a broadening of the energy levels. Additionally, the dependence of the activation gap on the magnetic field changes. Whereas it is proportional to $\sqrt{B}$ in pristine graphene, we rather observe a linear dependence on $B$ in hydrogenated graphene. This indicates a change from a linear dispersion near K in the band structure of pristine graphene to a parabolic one for hydrogenated graphene \cite{goerbig2009quantum}.

The differences between the band structure of pristine and hydrogenated graphene are pictured in Fig. \ref{fig:QHE}d with the dispersion relation around the Fermi level shown in the left panels and the density of states in a magnetic field in the right panels. 
As illustrated in the figure, the dispersion around the $K$-point changes from gapless-linear to quadratic with a small gap, the distance between the Landau levels decreases considerably and the levels broaden. Note that the image is not scaled.

From the magnetic field-dependence of the activation gap the effective electron mass can be calculated using that

\begin{equation}
    \Delta_{\mathrm{LL}} = \Gamma + 2\Delta_\mathrm{a} = \frac{\hbar eB}{m^*}.
\end{equation}

Here $\Delta_{\mathrm{LL}}$ is the distance between two Landau levels, which is the sum of the broadening $\Gamma$ and twice the activation gap $\Delta_\mathrm{a}$. $\hbar$ is the Planck constant, $e$ the electron charge and $B$ the magnetic field. The effective electron mass $m^*$ can then be calculated from the slope in Fig. \ref{fig:QHE}c and is found to be $m^*$ = 0.24$\pm$0.04 $m_\mathrm{e}$ for $\nu$=-2 after 25 minutes of plasma exposure and $m^*$ = 0.4$\pm$0.1 $m_\mathrm{e}$ after 30 minutes of plasma exposure.

\begin{figure}
    \centering
    \includegraphics[width=\linewidth]{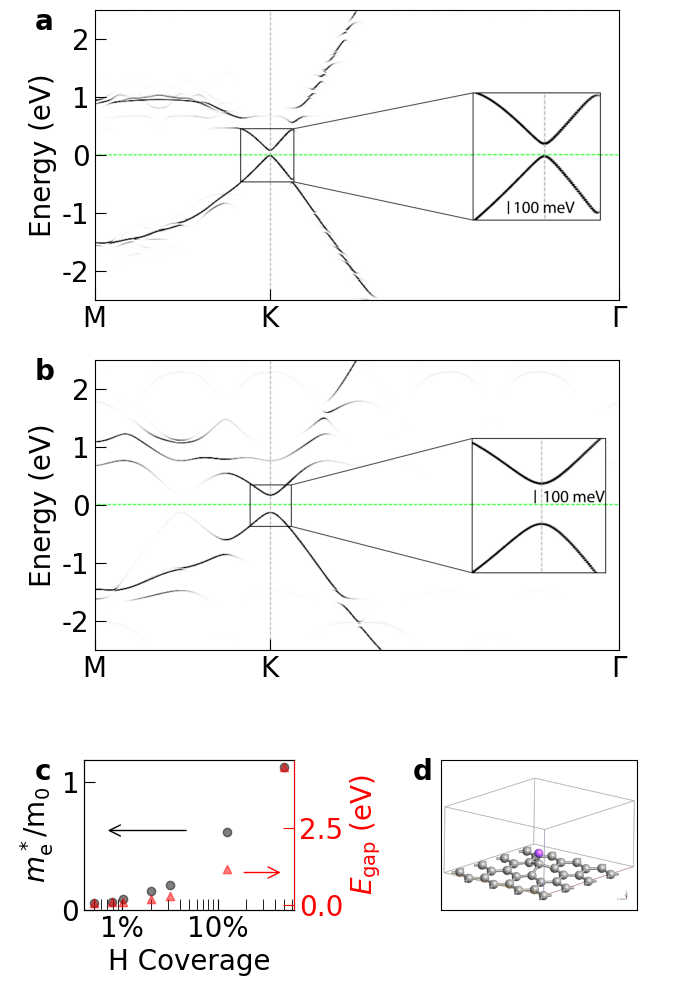}
    \caption{a. Band structure for a graphene sheet consisting of 128 carbon atoms with a single hydrogen atom attached to it.  b. Band structure for a graphene sheet consisting of 32 carbon atoms with a single hydrogen atom attached to it. c. Effective electron mass (gray dots) as function of hydrogen coverage, extracted from simulations. The effective electron mass was calculated using the second derivative of the conduction band near the K point. On the right axis, the energy gap is plotted (red triangles). d. Setup of the simulation. A base cell of in this case 32 carbon atoms is used, to which a single hydrogen atom (purple) is added. The hydrogen is bonded to one of the carbon atoms. Forcefield optimisation induces buckling of the graphene sheet \cite{reaxFF_CH_2017}.}
    \label{fig:coverage}
\end{figure}

\subsection{Band structure Simulations}

To help interpreting the changes in the band structure of graphene due to hydrogen plasma exposure, we have also performed simple band structure simulations using the Synopsys QuantumATK software \cite{QuantumATK_referencepaper}. We used a base cell of varying size  with periodic boundary conditions and added one hydrogen atom per base cell to the graphene lattice, which allows us to vary the relative hydrogen coverage of the graphene. Fig.~\ref{fig:coverage}d shows the base cell for the case of 32 carbon atoms. Although the simulation has been done on a 3D system, we can extract the 2D behaviour by fixing the distance in the perpendicular direction between two layers to 6.709 \AA. At such a large interlayer distance any effects of interlayer interactions can be safely neglected.
The geometry was optimized by minimising the forces on each atom using the ReaxFF potential \cite{reaxFF_CH_2017}. The Slater-Koster tight-binding model was used to simulate the band structure \cite{PhysRevB.58.7260,PhysRevB.82.075420}. 

 Fig.~\ref{fig:coverage}a and b show the effective band structure of a cell consisting of 128 and 32 C atoms respectively, each with a single H atom bonded to them. The effective band structure is constructed by unfolding the band structure of the cell into the Brillouin zone of the graphene base cell \cite{PhysRevLett.104.236403}. Consistent with our measurements, addition of hydrogen to the graphene sheet opens up a band gap, which increases with coverage. The relation can be seen in Fig.~\ref{fig:coverage}c. The bands also become quadratic near the $K$-point and from the curvature of the bands we can determine the effective electron mass using  

 \begin{equation}
     m^* = \frac{1}{\hbar^2} \frac{\partial^2E}{\partial k^2}.
 \end{equation}

Here $\hbar$ is the reduced Planck constant, $E$ is the energy and $k$ the momentum.

We have also done simulations of the band structure using deuterium atoms instead of hydrogen atoms. As these are not significantly different from the hydrogen simulations, we can conclude that tritium will likely behave similarly.

Fig.~\ref{fig:coverage}c shows the simulated band gap as a function of hydrogen coverage, which is in reasonable agreement with recent literature \cite{ValentinaTozzini, li2015structural}. 

The calculated effective electron mass as a function of hydrogen coverage is also shown in Fig.~\ref{fig:coverage}c. When comparing this to our experimentally determined masses of 0.24$\pm$0.04 m$_\mathrm{e}$ and 0.4$\pm$0.1 m$_\mathrm{e}$, we can deduce a hydrogen coverage of 4$\pm$1\% and 7$\pm$2\% respectively for the states shown in Fig.~\ref{fig:QHE}c.


\section{Conclusion and Outlook}

In conclusion, we have exposed graphene samples to hydrogen plasma and shown using Raman spectroscopy and transport measurements that this causes graphene hydrogenation. We have shown that hydrogenation causes a band gap to open in the graphene  and that the Landau level distance decreases significantly after hydrogenation. At $\approx 4\%$ hydrogen coverage the Landau level distance already depends linearly on magnetic field, indicating  that the band structure near the Fermi energy changes from a linear to a quadratic dispersion relation.

\begin{acknowledgments}
This publication is part of the project “One second after the Big Bang” NWA.1292.19.231 funded by the Dutch Research Council (NWO). It was supported by HFML-FELIX/NWO-I, member of the European Magnetic Field Laboratory (EMFL). C.S.A. Müller is grateful to Prof. F.A. Zwanenburg for his support. The authors gratefully acknowledge the MESA+ NanoLab staff and facilities for their support in making the device fabrication possible. We would like to thank V. Tozzini for useful discussions on the band stucture simulations and her insightful comments on the paper. We are grateful to the PTOLEMY Collaboration for the continuous exchange of ideas.

\end{acknowledgments}





%

\end{document}